\documentclass[a4paper,conference]{IEEEtran}
\usepackage[a4paper, left=0.514in, right=0.514in, bottom=1.569in, top=0.75in]{geometry}
\usepackage{graphicx}
\usepackage{amsmath}
\usepackage{adjustbox}
\usepackage{amssymb}
\usepackage{booktabs}
\usepackage{caption}
\usepackage{subcaption}
\usepackage{multirow}
\usepackage{amsfonts}
\usepackage{algorithm}
\usepackage{algorithmic}
\usepackage{hyperref}
\usepackage{xcolor}

\ifCLASSINFOpdf
\else
\fi
\begin{document}
%
\title{Semi-supervised Conditional GAN for Simultaneous Generation and Detection of Phishing URLs: A Game theoretic Perspective}


\author{\IEEEauthorblockN{Sharif Amit Kamran\IEEEauthorrefmark{1}, Shamik Sengupta\IEEEauthorrefmark{2} 
and Alireza Tavakkoli\IEEEauthorrefmark{3}}
\IEEEauthorblockA{
\textit{University of Nevada, Reno}
NV, USA\\
skamran@nevada.unr.edu\IEEEauthorrefmark{1}, ssengupta@unr.edu\IEEEauthorrefmark{2}, 
tavakkol@unr.edu\IEEEauthorrefmark{3}}
}

\maketitle

\begin{abstract}
Spear Phishing is a type of cyber-attack where the attacker sends hyperlinks through email on well-researched targets. The objective is to obtain sensitive information by imitating oneself as a trustworthy website. In recent times, deep learning has become the standard for defending against such attacks. However, these architectures were designed with only defense in mind. Moreover, the attacker's perspective and motivation are absent while creating such models. To address this, we need a game-theoretic approach to understand the perspective of the attacker (Hacker) and the defender (Phishing URL detector). We propose a Conditional Generative Adversarial Network with novel training strategy for real-time phishing URL detection. Additionally, we train our architecture in a semi-supervised manner to distinguish between adversarial and real examples, along with detecting malicious and benign URLs. We also design two games between the attacker and defender in training and deployment settings by utilizing the game-theoretic perspective. Our experiments confirm that the proposed architecture surpasses recent state-of-the-art architectures for phishing URLs detection. 
\end{abstract}


%

\section{Introduction}
Historically, cyber-security companies have used machine learning as a defensive measure to identify malicious websites and suspicious network activity. Given the massive surge in phishing attacks in recent times, it propelled companies to advertise novel systems for detecting inbound threats. According to the FBI \cite{fbi}, phishing was the most common type of cybercrime in 2020 and phishing incidents nearly doubled from 114,702 incidents in 2019 to 241,324 incidents in 2020. Out of this, 65\% of active hacker groups relied on spear-phishing to carry out the primary infection vector as reported in \cite{symantec}. In another report by Verizon \cite{verizon}, the attackers used email as a carrier for 96\% of the targeted attacks. While sites such as Phishtank \cite{phishtank} exist for sharing and verifying plausible malicious or phishing URLs, there has been little exploration into how attackers can also utilize machine learning to their advantage for generating such URLs.

The necessity for efficient countermeasures has made spear phishing a popular field of research in recent times. Effectively, two main types of methods for phishing URL detection have emerged: (i) Database of Blacklists and Whitelists URLs \cite{jain2016novel}, (ii) Character or feature level URL detection \cite{tajaddodianfar2020texception,seymour2018generative}. The database approach is quite inefficient due to not being updated on time or overlooking possible attacks from new URLs. The second approach is quite convenient for possible new attacks due to extracting latent representation of the feature level information and utilizing that to train a deep learning models (RNN \cite{huang2019phishing}, LSTM \cite{su2020research}, CNN \cite{aksu2019phishing}, GAN \cite{9644739}) to make the prediction. Compared to character-level detectors, the feature-level detection system suffers from out-of-distribution predictions because the features are constrained. As a result, the current state-of-the-art approaches utilize deep learning to learn from distinct characters of URLs to improve detection performance and enable zero-day phishing defense \cite{tajaddodianfar2020texception,huang2019phishing,aleroud2020bypassing}.

Cyber-security companies design this system with only defense in mind. The underlying motivation and perspective of the attacker are not taken into account, so the system can't formulate the best defense strategy for current or future attacks. We need a game-theoretic approach to understand the rational decision-making process of the attacker and how the defender should counter-play in this game. By taking all these into we propose the following in this paper: 

\begin{itemize}
  \item We propose a Conditional Generative Adversarial Network with novel training strategy for simultaneous generation and detection of phishing URL,  trained in a semi-supervised manner.
  \item We design two novel games based on game theory, i) between the pseudo attacker and the discriminator for training our conditional GAN, and ii) between the real hacker and the defender (the discriminator) for deployment in the wild.
  \item Our model outperforms other state-of-the-art architectures by learning from real and adversarial phishing URLs.
  \item Our discriminator detects URLs in real-time and is robust against adversarial examples.
\end{itemize}

\begin{figure*}[htp]
    \centering
    \includegraphics[width=\linewidth]{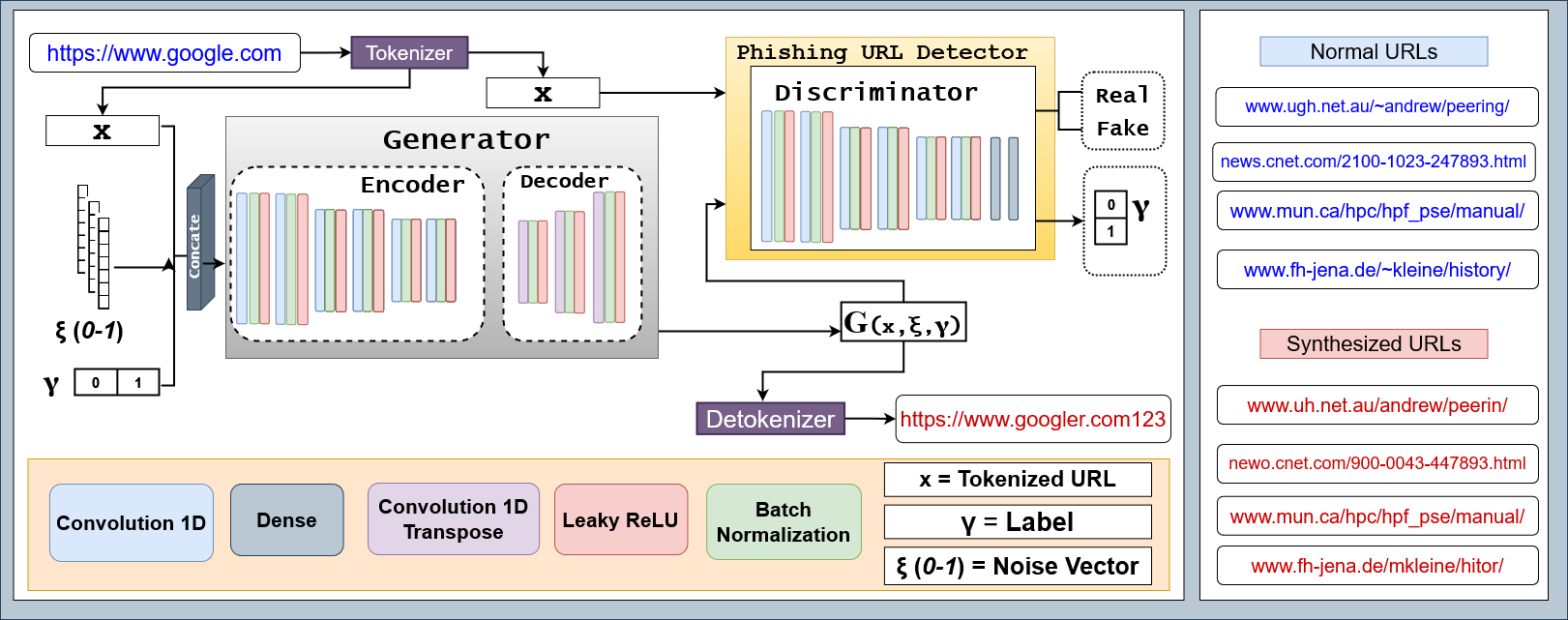}
    \caption{Proposed Conditonal GAN consists of a single Generator and Discriminator where the Generator takes the Real Benign or Malicious URL, a noise vector, and the class labels as input. The Generator outputs a synthesized URL, whereas the Discriminator categorizes Benign or Malicious URL, and detects adversarial examples.}
    \label{fig1}
\end{figure*}
\section{Literature Review}

Over the years, cyber-security firms have extensively utilized Machine Learning for detecting Phishing URLs and malicious sites. These works ranged from SVM \cite{zouina2017novel,jain2018phish,niu2017phishing}, Random Forest \cite{subasi2017intelligent,jagadeesan2018url}, KNN \cite{suleman2019optimization,xiao2020malicious} and a combinations of this techniques\cite{altaher2017phishing,pandey2018identification}. After the advent of Deep Learning, Convolutional Neural Networks (CNN) and Recurrent Neural Networks (RNN) have become the norm for automated feature extraction from a huge volume of data without any human intervention. Some recent works involving deep learning are based on CNN \cite{yi2018web,yang2019phishing,aksu2019phishing}, CNN with Multi-headed Attention \cite{xiao2020cnn}, RNN with Attention module \cite{huang2019phishing}, and LSTM \cite{chen2018phishing,su2020research,liang2019bidirectional}. There are two problems with this technique, a) the models are trained on lexical features such as URL Length, Count of top-level-domain, Number of punctuation symbols, and b) There are no adversarial elements present in training strategies, meaning no synthetic samples are generated or utilized for predictions.  The first problem can be addressed by training the model on character-level features rather than lexical or word-level features. On the other hand, the second problem can be solved by utilizing Generative Adversarial Networks for training the model for both real and adversarial examples.

With the popularity of Generative adversarial networks (GAN), there has been a surge in NLP applications ranging from text generation \cite{zhang2017adversarial,haidar2019textkd,fedus2018maskgan}, language models, \cite{xu2018dp,wang2018topic} and text classification \cite{croce2020gan,li2018learning}. The reason being, GANs can extract and learn fine-grained information from texts and utilize that to create synthetic examples using a generator architecture. Along with that, it utilizes a discriminator architecture for distinguishing between real and synthetic examples. By leveraging the generator and discriminator, the GAN can be trained in an adversarial setting with multiple cost functions and weights. Generative networks has also been used for synthesizing and identifying malicious URLs from lexical \cite{aleroud2020bypassing,robic2019detection,shirazi2020improved} or character-level data  \cite{anand2018phishing,tajaddodianfar2020texception}. The latter work involves training the generator to synthesize each URL character by sampling from a random noise vector, and the architecture is called unconditional GAN. This type of GAN suffers from "mode collapse" \cite{bau2019seeing,srivastava2017veegan} due to sampling from only a noise vector. As a result, the generator creates garbage data that the discriminator fails to distinguish, leading to nonconvergent adversarial loss. In contrast, conditional GAN incorporates both original data (malicious or benign URL) and label with the noise vector as input. As a result, it improves the overall model training and helps to synthesize more realistic output. 

Another major problem with the current state-of-the-art architecture is that they heavily emphasize defensive measures, somewhat ignoring the attacker's perspective in this scenario. A cyber-defense scenario almost always depends on game theory to understand the attacker's motives and perspective for maximizing the defender's reward. Many similar games have been designed for phishing URL detection with humans in the loop \cite{nwobi2011game,tchakounte2020game,figueroa2017adversarial} and adversarial games for generative networks \cite{jin2020multi,oliehoek2017gangs,hsieh2019finding}.  However, no work has been proposed, combining adversarial components of GAN and game-theoretic perspectives of attacker-defender for detecting phishing URLs. To alleviate this, we propose a novel conditional GAN that detects phishing URLs and is robust against adversarial examples. Furthermore, we design two games, one for training our architecture and the other for deploying it to a real environment for better understanding the attacker's perspective. Our quantitative result proves that the proposed technique surpasses state-of-the-architectures and synthesizes realistic benign and malicious URLs.

\section{The Proposed Methodology}
We propose a conditional generative adversarial network (GAN) comprising a generator for producing adversarial phishing URLs and a discriminator for robust classification of malicious/benign URLs. We also introduce two non-cooperative and non-zero-sum games between the i) Pseudo Attacker \& Discriminator and ii) Attacker  \& the Phishing URL Detector. By formalizing the game concept and addressing the limitations of current systems, we illustrate how generative networks can produce robust results in Phishing URLs detection and synthesis tasks.
First, we elaborate on the proposed architecture in Sections \ref{gen}, \ref{disc}, \ref{wet}. Next, we structure the two games based on players, their utilities, and action spaces in Section \ref{game1}, \ref{game2}.

\subsection{Generator}
\label{gen}
Combining generative networks with an auxiliary classification module has been shown to produce realistic text-generation and accurate class prediction as observed in \cite{croce2020gan,stanton2019gans}. We adopt a similar architecture to Auxiliary Classification GAN \cite{odena2017conditional} by incorporating a class conditioned generator and a multi-headed discriminator for categorical classification and adversarial example recognition as illustrated in Fig.~\ref{fig1}. The generator concatenates the original URL $x$, label $y$, and a noise vector $\zeta$ as input and generates $G(x,\zeta,y)$ malicious/benign URL. The noise vector, $z$, is smoothed with a Gaussian filter with $\sigma=3$ before pushing as input.  The generators consist of Convolution, Transposed Convolution, Batch-Normalization, and Leaky-ReLU activation layers. We use convolution for downsampling $3\times$ and use transposed convolution to upsample $3\times$ again to make the output resolution the same as the input. For regular convolution we use kernel size, $k=3$, stride, $s=1$ and padding, $p=1$. For downsampling convolution and transposed convolution we use kernel size, $k=3$, stride, $s=2$ and padding, $p=1$. Convolution and Transposed Convolutions are followed by Batch-normalization and Leaky-ReLU layers. The encoder consists of six convolution layers. The number of features are $[F1,F2,F3,F4,F5,F6]$= $[32,32,64,64,128,128]$. The decoder consists of three transposed convolution layers, and they have $[D1, D2, D3]$ =$[128,64,32]$ number of features.  The generator has an input and output dimension of $200\times67$ and incorporates Sigmoid activation as output. 

\subsection{Discriminator}
\label{disc}
In contrast, the discriminator takes both real URL $x$ and adversarial URL $G(x,\zeta,y)$ as input sequentially. It simultaneously predicts if the example is real or adversarial and classifies the signal as either Benign or  Malicious. The discriminator consists of Convolution, Batch-Normalization, Leaky-ReLU activation, and Dense layers. We use convolution for downsampling $3$ times. For convolution we use kernel size, $k=3$, stride, $s=1$ and padding, $p=1$, except for downsampling convolution where we use stride, $s=2$. The convolution layer is succeeded by Batch-normalization and Leaky-ReLU layers. After that, we use two fully connected layers.  The encoder consists of six convolution layers and two dense layers. The number of features are for each layer is $[F1,F2,F3,F4,F5,F6]$= $[16,16,32,32,128,128,256,64]$. We use two output activation: classification with Softmax activation for Benign/Malicious classes and Sigmoid activation for real/adversarial example detection.

\subsection{Weighted Cost Function and Adversarial Loss}
\label{wet}
We use LSGAN \cite{mao2017least} for calculating the adversarial loss and training our Genarative Network. The objective function for our conditional GAN is given in Eq.~\ref{eq1}. 
\begin{multline}
    \mathcal{L}_{adv}(G,D) =  \mathbb{E}_{x,y} \big[\ (D(x,y) -1)^2 \big]\ + \\ \mathbb{E}_{x,y} \big[\ (D(G(x,\zeta,y),y)+1))^2 \big]\
\label{eq1}
\end{multline}
In Eq.~\ref{eq1}, we first train the discriminators on the real URLs, $x$. After that, we train with the synthesized URLs, $G(x,\zeta,y)$. We begin by batch-wise training the discriminators $D$ on the training data for couple of iteration. Following that, we train the $G$ while keeping the weights of the discriminators frozen. For classification of Benign and Malicious URLs, we use categorical cross-entropy as in Eq.~\ref{eq2}.
\begin{equation}
    \mathcal{L}_{class}(D) = -\sum^{k}_{i=0} y_i\log(y'_i)
\label{eq2}
\end{equation}
The generators also incorporate the reconstruction loss (Mean Squared Error) as shown in Eq.~\ref{eq3}. By incorporating the loss, we ensure the synthesized URLs as convincing as the real URLs.

\begin{equation}
    \mathcal{L}_{rec}(G) = \mathbb{E}_{x,y} \Vert G(x,\zeta,y) - x \Vert^2
    \label{eq3}
\end{equation}

By incorporating Eq.~\ref{eq1}, \ref{eq2}  we can formulate our final objective function as given in Eq.~\ref{eq4}.
\begin{equation}
\min \limits_{G,D} \big( \max \limits_{D}  (\lambda_{adv} \mathcal{L}_{adv}(D)) +  \lambda_{rec}\ \mathcal{L}_{rec}(G) + \lambda_{class} \mathcal{L}_{class}(D) \big)
\label{eq4}
\end{equation}
Here, $\lambda_{adv}$, $\lambda_{rec}$, and $\lambda_{class}$ implies different weights, that is multiplied with their respective losses. The loss weighting decides which architecture to prioritize while training. For our system, more weight is given to the $\mathcal{L}_{class}(D)$ for better categorical prediction, and thus we select bigger $\lambda$ value for this. 

\subsection{Game during Model Training}
\label{game1}
We begin by contextualizing a novel two-person game similar to \cite{katzef2020distributed}, using the proposed conditional GAN between the Pseudo Attacker and the Phishing URL detector. Both the players are interlocked in a non-cooperative, non-zero-sum  game signified formally by the following tuple of players, action spaces, and utility variables: $Game_{MT}= \{P, D\}, \{\Theta_G, \Theta_D\}, \{u_G, u_D\}$ where

\begin{itemize}
	\item P is the pseudo-attacker who controls generator G by utilizing its $\theta_G$-parametrized function approximator, $G(x,\zeta,y)$ with $G : {\Bbb R} \rightarrow  {\Bbb R}^d$, to choose its actions,
	\item D is the phishing URL detector, which uses a $\theta_D$-parametrized function approximator, $D(G(x),y)$ with $D : {\Bbb R}^d\rightarrow  {\Bbb R}$, to choose its actions,
	\item $\theta_G$ is the action space of the generator, where $\theta_G \in \Theta_G$,
	\item $\theta_D$ is the action space of the discriminator, where $\theta_D \in \Theta_D$,
	\item $u_G$ and $u_D$ are the generator and discriminator's utility variables.
\end{itemize}

Based on the assumptions above, we can rewrite the Eq.~\ref{eq1} as Eq.~\ref{eq5}. Here $p(x)$ is the distribution of the given data samples, $x \in  {\Bbb R}$, $p(\zeta)$ is the the Gaussian distribution of the noise sample, $\zeta \in [0,1]$ and $p(y)$ is the distribution of the ground truth, $y\in \{0,1\}$.

\begin{multline}
    u_{adv}(\theta_G,\theta_D) =  \mathbb{E}_{x\sim p(x),y\sim p(y)} \big[\ D(x,y) -1 \big]\ ^2 + \\ \mathbb{E}_{x\sim p(x),\zeta\sim p(\zeta),y\sim p(y)} \big[\ D(G(x,\zeta,y),y)+1) \big]\ ^2
\label{eq5}
\end{multline}

The phishing URL detector can also distinguishes between "Benign" and "Malicious" URL using categorical-cross entropy loss in Eq.~\ref{eq2}. Moreover, the generator also tries to fool the discriminator with realistic sample using L2 bounded reconstruction loss in Eq.~\ref{eq3}. We can rewrite Eq,~\ref{eq2} and Eq.~\ref{eq3}, and formulate it into, Eq.~\ref{eq6}, and Eq.~\ref{eq7}. By combining Eq.~\ref{eq5}, Eq.~\ref{eq6}, Eq.~\ref{eq7}, we can find the final utility function given in Eq.~\ref{eq8}. Here, $\lambda$ is the weight values of each utility functions similar to weight values in Eq.~\ref{eq4}.

\begin{equation}
    u_{class}(\theta_G,\theta_D) =   \mathbb{E}_{y\sim p(y),y'\sim p(y')} \big[\ -\sum^{k}_{i=0} y_i\log(y'_i) \big]\
\label{eq6}
\end{equation}

\begin{equation}
    u_{rec}(\theta_G,\theta_D) =   \mathbb{E}_{x\sim p(x),\zeta\sim p(\zeta),y\sim p(y)} \big[\ \Vert G(x,\zeta,y) - x \Vert^2 \big]\
\label{eq7}
\end{equation}

\begin{multline}
    u(\theta_G,\theta_D) =    \lambda_{adv} \big[u_{adv}(\theta_G,\theta_D) \big]\ +  \lambda_{rec} \big[\ u_{rec}(\theta_G,\theta_D) \big]\ + \\ \lambda_{class}\big[u_{class}(\theta_G,\theta_D) \big]\
\label{eq8}
\end{multline}

\begin{figure}[!t]
    \centering
    \includegraphics[width=\linewidth]{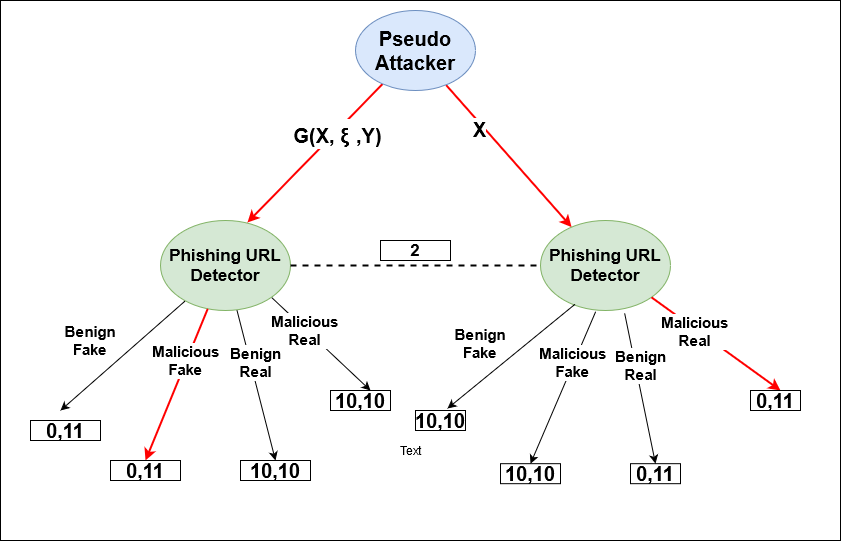}
    \caption{Game during Model training between the Pseudo Attacker and Phishing URL detector. The game has eight possible outcomes, based on the actions taken by the Attacker and Defender. }
    \label{fig2}
\end{figure}

Next, we design a Kuhn tree between the Pseudo Attacker (Attacker) and the Phishing URL detector (Defender) as illustrating in Fig.~\ref{fig2}. We assume the strategy space of the Pseudo Attacker consists of two actions: providing (i) adversarial example $G(x,\zeta,y)$ and (ii) real example $x$. In contrast, the Phishing URL detector can take four actions: classify the sample as (i) Benign Fake, (ii) Malicious Fake, (ii) Benign Real, or (iv) Malicious Real. We further assume that the Pseudo Attacker will make the first move, and then the  Phishing URL detector will follow by taking one of the four possible actions. It should also be noted that the Phishing URL detector does not have information about the sample's type and if they are real or fake.

The game has eight possible outcomes, based on the actions taken by the Attacker and Defender. Suppose the attacker uses the adversarial example $G(x,\zeta,y)$, then the defender can use any of the four actions to classify it. We can see in Fig.~\ref{fig2}, the best strategy for the defender would be to classify it as either Malicious Fake or Benign Fake. On the other hand, if the attacker uses the real example $x$, then the defender's best strategy would be to classify it as either Malicious Real or Benign Real. The payoff for the attacker and the defender were created using the $\lambda$ values in Eq.~\ref{eq8}. The maximum profit the attacker can gain by fooling the Phishing URL detector (discriminator) can be $\lambda_{rec} = 10$. The minimum can be 0.  Contrarily, the defender would get a maximum value of 11, by summing $\lambda_{cls} = 10$ and $\lambda_{adv}=1$. The minimum payoff would be 10, because of the correct class prediction.

\begin{figure}[!t]
    \centering
    \includegraphics[width=1\linewidth]{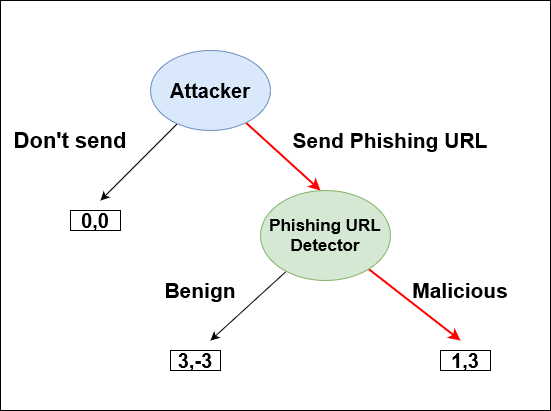}
    \caption{Game after Model Deployment between the Real Attacker and Phishing URL detector.}\label{fig3}
\end{figure}
\subsection{Game after Model Deployment}
\label{game2}
Next, we design a novel two-person game for model deployment, which would be played between the real-life attacker and the Phishing URL detector (Defender) in a non-cooperative, non-zero-sum  game signified formally by the following tuple of players, action spaces, and utility variables: $Game_{MD} = \{A, D\}, \{\Theta_A, \Theta_D\}, \{u_A, u_D\}$ where

\begin{itemize}
	\item A is the attacker who either sends and doesn't send the phishing URL and the action is defined by $\theta_A \in \{ \text{Dont Send, Send} \} $,
	\item D is the phishing URL detector, which either classifies the URL as Benign or Malicious and the actions is defined by $\theta_D \in \{ \text{Benign, Malicious} \} $,
	\item $\theta_A$ is the action space of the attacker, where $\theta_A  \in \Theta_A$.
	\item $\theta_D$ is the action space of the discriminator, where $\theta_D \in \Theta_D$,
	\item $u_A$ and $u_D$ are the attacker and phishing URL detector's utility variables.
\end{itemize}

Based on this above assumption we create our Kuhn tree for Model Deployment and is illustrated in Fig.~\ref{fig3}. This time the strategy space of the attacker consists of two actions: either to (i) Send Phishing URL or (ii) Don't Send. Contrarily, the Phishing URL detector can take two actions: classify the sample as (i) Benign or (ii) Malicious. The assumption also remains the same: the attacker will make the first move, and then the  Phishing URL detector will follow by taking one of the two possible actions.  The utility values are such that $u_A \in {\Bbb R}$ and $u_D \in {\Bbb R}$. 

The game has three possible outcomes, based on the actions taken by the attacker and defender. If the attacker does not send the phishing URL, they will both get a payoff of 0. On the other hand, if the attacker sends the phishing URL, the defender either has to classify it as malicious or benign. We can see in Fig.~\ref{fig3}, the best strategy for the defender would be to classify it as Malicious. The payoff for the attacker and the defender were created using arbitrary values. The maximum profit the attacker can gain carrying out a successful attack is 3. The minimum can be 1, if the attacker is unsuccessful.  In contrast, the defender would get a maximum value of 3 by classifying it as malicious. The minimum payoff would be -3, because it would mean losing all the sensitive information.

\section{Experiments}

In the next section, we detail our model experiments and evaluate our architecture based on quantitative metrics. First, we elaborate on the structuring and pre-processing of our dataset in Sec.~\ref{subsec:dataset}. Then detail our hyper-parameter selection and tuning in Sec.~\ref{subsec:hyper}. Next, we describe our adversarial training scheme in Sec. ~\ref{subsec:training}. Also, we compare our architecture with existing state-of-the-art models based on some quantitative evaluation metrics in Sec.~\ref{subsec:quant}. 

\subsection{Dataset}
\label{subsec:dataset}
We used the publicly curated and standardized dataset provided in Kaggle \cite{kaggle}. It consists of 500,000+ benign and malicious URLs collected from various sources. Next, we sample 50,000 benign and malicious URLs with a max length of 200 characters.  The dataset includes an equal portion of benign and malicious samples. We use a dictionary of 67 unique characters (twenty-six alphabets, ten digits, and thirty-one special characters) to convert the data into a one-hot encoding matrix. Now, each URL ends up being a 2-D matrix of size $200\times67$. We use 40,000 URLs for training and 10,000 as test-set. For training, we use 5-fold cross-validation and choose the best model for testing.
\subsection{Hyper-parameter tuning}
\label{subsec:hyper}
For adversarial training, we used LS-GAN loss \cite{mao2017least}. We picked $ \lambda_{rec} =10$ and $ \lambda_{class} =10$ (Eq.~\ref{eq4}). For optimizer, we used Adam \cite{kingma2014adam}, with learning rate $\alpha=0.0002$, $\beta_1=0.5$ and $\beta_2=0.999$. We train with mini-batches with batch size, $b=64$ for 200 epochs. It took approximately 6 hours to train our model on NVIDIA GPU. {\color{blue}\href{https://github.com/SharifAmit/Semi-supervised-Phishing-Detection-GAN}{Code Repository Link.}} 

\begin{algorithm}[!tp]
\caption{Phishing URL detection Training}
\label{alg1}
\begin{algorithmic}[1]
 \renewcommand{\algorithmicrequire}{\textbf{Input:}}
 \renewcommand{\algorithmicensure}{\textbf{Output:}}
  \REQUIRE $x^{i} \epsilon X$, $y^i_{adv} \epsilon {Y_{adv}}$, $y^i_{cls} \epsilon {Y_{cls}}$
 \ENSURE $G$, $D$
  \STATE \textbf{Initialize hyper-parameters}: \\$max\_epoch$, $b$, $max\ d\_iter$, $\omega_{D}$, $\omega_{G}$, $\alpha_{D}$,  $\alpha_{G}$, $\beta_{D}$, $\beta_{G}$, $\lambda_{adv}$, $\lambda_{rec}$, $\lambda_{cls}$
  \FOR{$e=0\ to\ max\_epoch$}
  \STATE Sample $x,y_{adv},y_{cls}$, using batch-size $b$ 
    \FOR{$d\_iter=0\ to\ max\ d\_iter$}
        \STATE $\mathcal{L}_{real}(D), \mathcal{L}_{cls}(D)\gets D(x,y_{adv},y_{cls})$  
        \STATE $\mathcal{L}_{fake}(D), \mathcal{L}_{cls}(D) \gets D(G_{c}(x),y_{adv},y_{cls})$
        \STATE $\mathcal{L}_{adv}(D) \gets \mathcal{L}_{real}(D) +\mathcal{L}_{fake}(D)$
        \STATE $\omega_{D} \gets \omega_{D}+ Adam(D,G,\omega_{D},\alpha_{D},\beta_{D})$
    \ENDFOR
    \item[] \textbf{Freeze $\omega_{D}$}
    \STATE Sample $x,y_{adv},y_{cls}$, using batch-size $b$ 
    \STATE $\mathcal{L}_{rec}(G_c) \gets G(x,\zeta,y), x$
    \STATE $\omega^c_{G} \gets \omega_{G}+ Adam(G,\omega_{G},\alpha_{G},\beta_{G})$
    \STATE $\mathcal{L}_{real}(D), \mathcal{L}_{cls}(D) \gets D(x,y_{adv},y_{cls})$  
    \STATE $\mathcal{L}_{fake}(D), \mathcal{L}_{cls}(D) \gets D(G_{c}(x),y_{adv},y_{cls})$
    \STATE $\mathcal{L}_{adv}(D) \gets \mathcal{L}_{real}(D) +\mathcal{L}_{fake}(D)$
    \STATE $\omega_{D} \gets \omega_{D}+ Adam(D,G,\omega_{D},\alpha_{D},\beta_{D})$
    \STATE Save weights and snapshot of $D,G$
    \STATE $\mathcal{L} \gets \lambda_{adv}(\mathcal{L}_{adv}) +  \lambda_{rec}(\mathcal{L}_{rec}) + \lambda_{cls}(\mathcal{L}_{cls}) $
  \ENDFOR
\end{algorithmic}
\end{algorithm}
\subsection{Training procedure}
\label{subsec:training}
In this section, we elaborate on our detailed algorithm provided in Algorithm~\ref{alg1}. To train our model, we start by initializing all the hyper-parameters. Next, we sample a batch of the real URLs $x$. We train the real URLs $D$. After that, we use $G$ to synthesize fake URLs  $G(x,\zeta,y)$ and use them for training our discriminator $D$ again. We train the discriminator in this manner for a couple of iterations. Following that, we calculate the adversarial loss, $\mathcal{L}_{adv}(D)$, the classifier loss, $\mathcal{L}_{cls}(D)$, and update the weights. We freeze the weights of the discriminator. Next, we train the generator and calculate the $\mathcal{L}_{rec}(G)$, and update the generator's weights. In the final stage, while keeping the discriminator's weights frozen, we jointly fine-tune the discriminator and generator together. We calculate the total loss by adding and multiplying with their relative weights. For testing, we save the snapshot of the model and its weights.

\subsection{Quantitative Evaluation}
\label{subsec:quant}
For finding the character similarity with the original URL, we benchmark synthesized adversarial URLs using four different metrics, i) Mean Squared Error (MSE), ii) Structural Similarity Index (SSIM), and iii) Normalized Mean Squared Error (NRMSE). Table.~\ref{table1} shows that SSIM for the test set has a 98.33\% score, which means the adversarial examples are structurally similar to the original signal. As for MSE and NRMSE, our model generates quite similar adversarial examples to real ones. It is important to note that we want to achieve lower MSE and RMSE. Similarly, we want to score higher for SSIM.

\begin{figure}[!tp]
    \centering
    \includegraphics[width=1\linewidth]{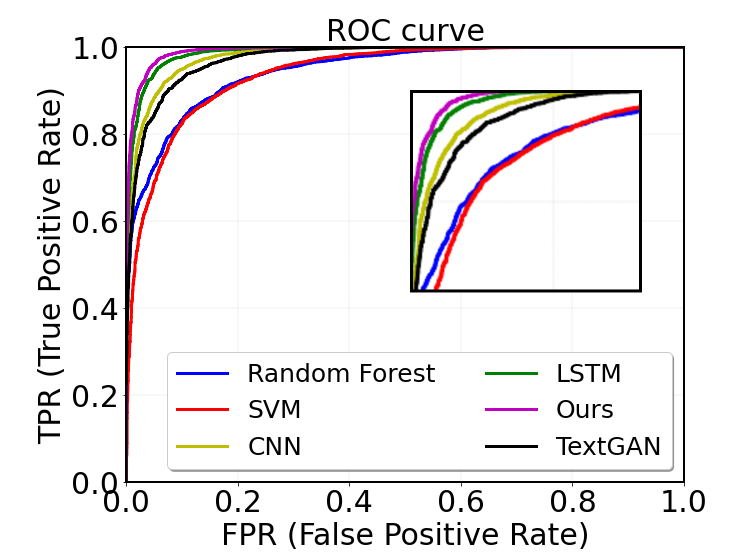}
    \caption{Our model outperforms other methods in terms of Area under the Receiver Operating Curve (ROC) for detecting Benign/Malicious Phishing URLs.}
    \label{fig4}
\end{figure} 

\begin{table}[!b]
\caption{\textbf{Generator's Performance}: Similarity between real and adversarial Phishing URLs.}
\centering
\begin{adjustbox}{width=0.8\linewidth}
\begin{tabular}{|c|c|c|c|}
\hline
 MSE & Structural Similarity  & Normalized RMSE\\ \hline
0.000579          &       0.9833      &  0.19708 \\ \hline
\end{tabular}
\end{adjustbox}
\label{table1}
\end{table}

For benign and malicious URL classification tasks, we compare our model with other state-of-the-art architectures as given in Table.~\ref{table2}. We further experiment on adversarial examples synthesized using our Generator from the above tests set, which we show in Table.~\ref{table3}. We use Accuracy (ACC), Sensitivity, Precision, F1-Score, and AUC (area-under-the-curve) for metrics. We can see for the first experiment, except for the architectures given in \cite{xiao2020cnn}, our model achieves the best score compared to other deep learning and machine learning derived architectures. The architecture in \cite{xiao2020cnn} uses 1D CNN and LSTM architectures. Because these architectures have millions of parameters, they might be overfitting on the data. Compared to them,  our architecture is quite lightweight and has around 81,000 parameters. Our model also has an inference speed of 0.64 ms (milisecond). Meaning it can process around 1500+ URLs in 1 second. The models in \cite{robic2019detection,tajaddodianfar2020texception} uses DC-GAN with character level data. However, they perform poorly compared to our model. We also provide AUC and illustrate all the model's Receiver operating curve (ROC) in Fig.~\ref{fig4}.

\begin{table}[!tp]
\caption{\textbf{Phishing URL classification} : Comparison of architectures trained and tested on \textbf{real} Phishing URLs}
\centering
\begin{adjustbox}{width=\linewidth}
\begin{tabular}{|c|c|c|c|c|c|}
\hline
 Method & ACC  & Sensitivity & Precision & F1-score & AUC\\ \hline
Proposed Method  &  \textbf{0.9552}   & \textbf{0.9600}    &  \textbf{0.9508}   & \textbf{0.9554}  & \textbf{0.9552}\\ \hline
TextGAN \cite{anand2018phishing} & 0.9135 & 0.9210 & 0.9073 & 0.9141 & 0.9135 \\ \hline
SVM \cite{yerima2020high} & 0.8638 & 0.8822 & 0.8508 &  0.8662 & 0.8638 \\ \hline
Random Forest \cite{shirazi2020improved} & 0.8688 & 0.8508 & 0.8825 & 0.8664 & 0.8688 \\ \hline
CNN \cite{xiao2020cnn} & 0.9251 & 0.9190 & 0.9303 & 0.9246 & 0.9251 \\ \hline
LSTM \cite{xiao2020cnn} & 0.9471 & 0.9522 & 0.9425 & 0.9473 & 0.9471 \\ \hline
\end{tabular}
\end{adjustbox}
\label{table2}
\end{table}

For our last experimentation,  we evaluate our discriminator and check its accuracy to detect adversarial examples as given in Table.~\ref{table3}. We use a test set with 50\% adversarial and 50\%  real data to carry out this benchmark. By combining the real and the adversarial examples synthesized by our generator, we can have this 50/50 test split. So from the test set of 10,000 URLs, we end up having 20,000 samples. We use Accuracy, Sensitivity, Specificity, F1-score, and AUC as standard metrics for measuring our model's performance. As shown in Table.~\ref{table3}, our model is quite robust against adversarial examples. However, the model performs better for detecting real examples compared to adversarial examples, which is validated by 100\% Specificity. Moreover, the model detects 74.9\% adversarial examples correctly, which is validated by the Recall. Consequently, it confirms the hypothesis that Generative Networks are far superior for detecting adversarial examples. 
\begin{table}[!t]
\caption{\textbf{Adversarial Example Detection}: Discriminator's performance for detecting adversarial examples }
\centering
\begin{adjustbox}{width=\linewidth}
\begin{tabular}{|c|c|c|c|c|c|}
\hline
 Method & ACC  & Sensitivity & Specificity & F1-score & AUC\\ \hline
Proposed Method  &  0.8745  & 0.749  & 1.000  & 0.856   & 0.8745 \\ \hline
\end{tabular}
\end{adjustbox}
\label{table3}
\end{table}
\section{Conclusion}
In this paper, we introduced a conditional Generative Adversarial Network with novel game-theoretic training strategy for simultaneous synthesis of adversarial examples and detecting Phishing URLs.  Our architecture outperforms previous techniques by adopting a semi-supervised learning scheme of synthesizing adversarial URLs while predicting their category. The model is best suited for real-time Phishing detection monitoring both locally (client-end) and globally (server-end), where it can perform robustly and effectively. One future work is to test the adversarial examples on other architectures and compare their robustness.

\bibliographystyle{IEEEtran}
\bibliography{reference}



\end{document}